\documentclass[iop]{emulateapj}
\usepackage{graphicx}
\usepackage{amsmath}

\def\Hubble{{\it Hubble }}

\def\Spitzer{{\it Spitzer }}

\def\Herschel{{\it Herschel }}
\def\Herschels{{\it Herschel}}
\def\Halpha{{\it $H_{\alpha}$}}
\def\O2{ [\ion{O}{2}]}
\def\micron{$\rm \mu$m }
\def\lirs{$L_{IR}$ }

\begin{document}

\title{De-Confusing blended field images using graphs and bayesian priors}

\author{Mohammadtaher Safarzadeh\altaffilmark{1},Henry C. Ferguson\altaffilmark{2}, Yu Lu\altaffilmark{3}, Hanae Inami\altaffilmark{4}, Rachel S. Somerville\altaffilmark{5}}

\altaffiltext{1}{Johns Hopkins University, Department of Physics and
Astronomy, 366 Bloomberg Center, 3400 N. Charles Street, Baltimore, MD
21218, USA,
\\Email: mts@pha.jhu.edu}
\altaffiltext{2}{Space Telescope Science Institute, 3700 San Martin Boulevard, Baltimore, MD 21218, USA}
\altaffiltext{3}{Kavli Institute for Particle Astrophysics and Cosmology, Stanford, CA 94309, USA}
\altaffiltext{4}{National Optical Astronomy Observatory, 950 North Cherry Avenue, Tucson, AZ 85719, USA}
\altaffiltext{5}{Department of Physics and Astronomy, Rutgers, The State University of New Jersey, 136 Frelinghuysen Road, Piscataway, NJ 08854, USA}

\begin{abstract}

We present a new technique for overcoming confusion noise in deep far-infrared \Herschel space telescope images making use of prior information from shorter $\lambda<2$\micron wavelengths. For the deepest images obtained by \Herschels, the flux limit due to source confusion is about a factor of three brighter than the flux limit due to instrumental noise and (smooth) sky background. 
We have investigated the possibility of de-confusing simulated \Herschel PACS-160\micron images by using strong Bayesian priors on the positions and weak priors on the flux of sources. We find the blended sources and group them together and simultaneously fit their fluxes. We derive the posterior probability distribution function of fluxes subject to these priors through Monte Carlo Markov Chain (MCMC) sampling by fitting the image. Assuming we can predict FIR flux of sources based on ultraviolet-optical part of their SEDs to within an order of magnitude, the simulations show that we can obtain reliable fluxes and uncertainties at least a factor of three fainter than the confusion noise limit of $3\sigma_{c} $=2.7 mJy in our simulated PACS-160 image. This technique could in principle be used to mitigate the effects of source confusion in any situation where one has prior information of positions and plausible fluxes of blended sources. For \Herschel, application of this technique will improve our ability to constrain the dust content in normal galaxies at high redshift.  
\end{abstract}

\begin{keywords}
{Noise:Confusion -- galaxies: Infrared -- galaxies: dust content -- Method: Markov Chain Monte Carlo}
\end{keywords}

\section{Introduction}
\label{sec:intro}

A significant fraction of the radiation emitted by stars and Active Galactic
Nuclei (AGN) over the lifetime of the universe is absorbed by dust and
re-radiated at long wavelengths. It is thus crucial to measure this re-emitted
radiation and develop an understanding of how dusty radiative transfer evolves in galaxies over cosmic time.
There has been a revolution over the past two
decades in our ability to measure far-infrared (FIR) radiation, with the
increasing sensitivity of sub-mm telescopes and detectors and with the launch of
the {\it Herschel} Far-infrared Observatory \citep{herschel}.  However, at
the flux levels relevant for typical galaxies at redshifts $z > 1$, our
deepest images of the sky at wavelengths 24-1000\micron are dominated by
confusion noise (Condon 1974). If we were to boost the Milky Way$'$s
star-formation rate (SFR) by a factor of 100, it still would not be detectable
in the deepest Herschel images if it were placed at redshifts $z > 2$.
Furthermore, the main-sequence of star-forming galaxies  \citep{elbaz_11, noeske_07}
drops below the confusion limit at $z>2$  \citep{magnelli_13}. It is therefore important to find techniques to overcome or mitigate
confusion noise.

Star-formation buried inside molecular clouds could be hidden from ultraviolet(UV)-optical observations due to 10-100 mag of extinction, and would be only revealed in the FIR. UV photons from the newly born stars are absorbed by dust (and for photons with $h\nu\!>\! 13.6$  eV by neutral hydrogen) inside the molecular cloud, and in return the absorbed energy by the dust is emitted in the FIR (see \citealt{silva_98,charlot_fall_00,Chevallard_13} for detailed modeling of  extinction in diffuse interstellar medium (ISM) and birth clouds).

At high redshifts, SFRs are inferred from an estimate of the unattenuated far-UV luminosity \citep[$L_{UV}$;][]{meurer_99,finkelstein_12,bouwens_12}, nebular line emissions \citep[such as \Halpha, Lyman-${\alpha}$,][]{moustakas_06}, rest frame 8 $\mu$m luminosity \citep{elbaz_11}, dust emission in the FIR \citep{magdis_10,heinis_14,wardlow_14} and Sub-mm observations \citep{blain_99}. A thorough review of this subject is presented in \citet{madau_dickinson_14}.

For galaxies with roughly constant SFR, the UV-continuum slope at $1200<\lambda<2000$ is approximately flat in $f_{\nu}$. For such galaxies, departures from a flat continuum are mostly due to dust. The dust attenuation is commonly estimated by measuring the UV-continuum slope $\beta$  \citep[$f_{\lambda} \propto \lambda^{\beta}$][]{meurer_99,calzetti_94}. Under the assumption that light missing from the UV is re-emitted in the FIR (i.e., ignoring scattering), local samples of galaxies have been used to calibrate the relation between the slope and the ratio of FIR to UV luminosity $IRX=\log{L_{IR}/L_{UV}}$\citep{meurer_99}. While this works reasonably well for normal star-forming galaxies, luminous infrared galaxies $(L_{IR}(8-1000 \mu m) > 10^{11} L_{\odot})$ which dominate the SFR density at ${\it z}\! >\!1$ (e.g. \citet{lefloch_05,caputi_07,magnelli_09,magnelli_11,murphy_11}) do not follow the suggested $\beta-IRX$ relation \citep{goldader_02}, suggesting hidden ongoing star formation activity in them. Moreover, the very large scatter in $\beta-IRX$ relation renders its application problematic \citep{conroy_13}.

Even for un-attenuated galaxies, the UV-continuum slope is a reliable measure of $L_{UV}$ only for galaxies with constant SFR. If a galaxy is experiencing quenching, the O and B stars die off and are not replaced by the next generation. Therefore, the old stellar population will dominate the UV-continuum making the slope redder (the same as dust would do) and skewing the underlying $\beta-IRX$ relation. This is an important issue because it is difficult to differentiate between an old stellar population and a dusty star forming galaxy (this is often referred to as the age--dust degeneracy).
 If the estimate of SFR based on the UV continuum slope is reliable, the correction that needs to be applied to obtain the unattenuated $L_{UV}$ is generally large--typically a factor of $\sim$5 for Lyman-break galaxies \citep{meurer_99,reddy_12}. Therefore relatively small photometric uncertainties of 10-20\% \citep[the quoted error on $\beta$ is 20\% in][]{adelberger_00,finkelstein_10} can balloon into uncertainties on the star-formation rates of 50-100\%.

Measurements in the FIR yield a more robust estimate of SFR but have their own biases \citep{kennicutt_09}. In the case of dwarf galaxies and metal poor systems, where the fraction of obscured star forming regions is low, the FIR does not trace the total SFR. To obtain the most reliable estimates of SFR, it is best to combine the estimates from UV and the FIR \citep{kennicutt_03}. It is important to note that FIR-based measurements of SFR are not a minor correction to the SFR estimates based on rest frame UV observations. \citet{reddy_12} found that 80\% of star formation in Lyman Break Galaxies (LBGs) at ${\it z}\sim 2$ is hidden in dust and is only revealed in FIR dust emission.

Currently studies of the massive end of the main sequence of star-forming galaxies at {\it z} $>$2 in the FIR have been limited to stacking analysis. Stacking has the intrinsic assumption that the underlying distribution of galaxies is a normal distribution where the mean and median are well behaved. Furthermore, the median or mean of some derived physical quantity for a sample of galaxies is not necessarily the same as the value for that physical quantity derived from the mean or median of the stacks \citep{ryan_14,vargas_14}. In stacking, the presence of local background around a particular object can skew the final results. Moreover, because galaxies are clustered, a portion of any detected signal could be due to physical neighbors. Detected sources are often subtracted before doing the stacking (to improve the S/N), but this risks subtracting a portion of the flux of the sources of interest. In addition, the estimate for the mean optical/FIR flux ratio from the stacks can be biased by sources in the tails of the distribution. Therefore individual detections (i.e. not stacked photometry) of high redshift galaxies will provide more reliable estimates of their SFRs; however source confusion restricts the ability to detect faint objects and de-blend neighbors. FIR studies of individual high redshift lensed galaxies are currently limited to a small number of systems \citep[e.g. ][]{egami_10}.

Confusion noise was first recognized in the context of deep  radioastronomical observations 
\citep{scheuer_1957}. An image can be considered as confusion limited when the uncertainties
in the measured fluxes of the sources are dominated by the uncertainties
due to overlapping sources.  \citet{murdoch_73} discussed two types of confusion: blending confusion which comes from a high number density of faint (but detectable if observed individually) sources in the beam and the latter photometric confusion which is due to sources with fluxes less than the instrumental detection limit ($S_{lim}$). The quadrature sum of these two
causes of photometric scatter is designated as confusion noise $\sigma_{conf}$.

Apart from radio astronomy \citep{blain_98,condon,condon_12}, confusion noise is important for many other types of observations as well, 
such as FIR observations \citep{spire_confusion,magnelli_13,galactic_cirrus,north_pole,cold_debris,wise_survey}, X-ray deep observations\citep{x_ray,barcon_92}, gravitational waves\citep{lisa}, weak lensing \citep{weak_lensing}, Sunyaev-Zel'dovich (SZ) cluster surveys\citep{sz}, high precision astrometry \citep{hogg_01} and studies of the galactic center\citep{sag_a_star,sag_a_star_stone}.
A typical rule of thumb has been that one reaches diminishing returns when there are more than 1/30 sources per beam. The classic treatment of source confusion assumes that you know nothing about the sources other than their statistical density on the sky. Moreover, it is assumed that the sources are distributed uniformly on the sky with Poisson statistics. However, we know galaxies are clustered and this changes the confusion noise limit \citep{takeuchi_04,negrello_04,barcon_92}. 

It is useful to consider an alternative hypothetical case, where one knows everything about all the sources in the image, even below the detection limit, except for one source. In this case, all of the contaminating sources can be subtracted perfectly. One is left with just the detector noise, residual Poisson noise from the subtracted sources and noise from the smooth sky background (cirrus and zodiacal dust in the case of FIR observations). The source in question is no longer contaminated by confusion. Provided there is not a systematic noise floor, the photometric uncertainty for this source will beat down as the square root of the exposure time. 

\begin{figure}
\includegraphics[scale=0.3]{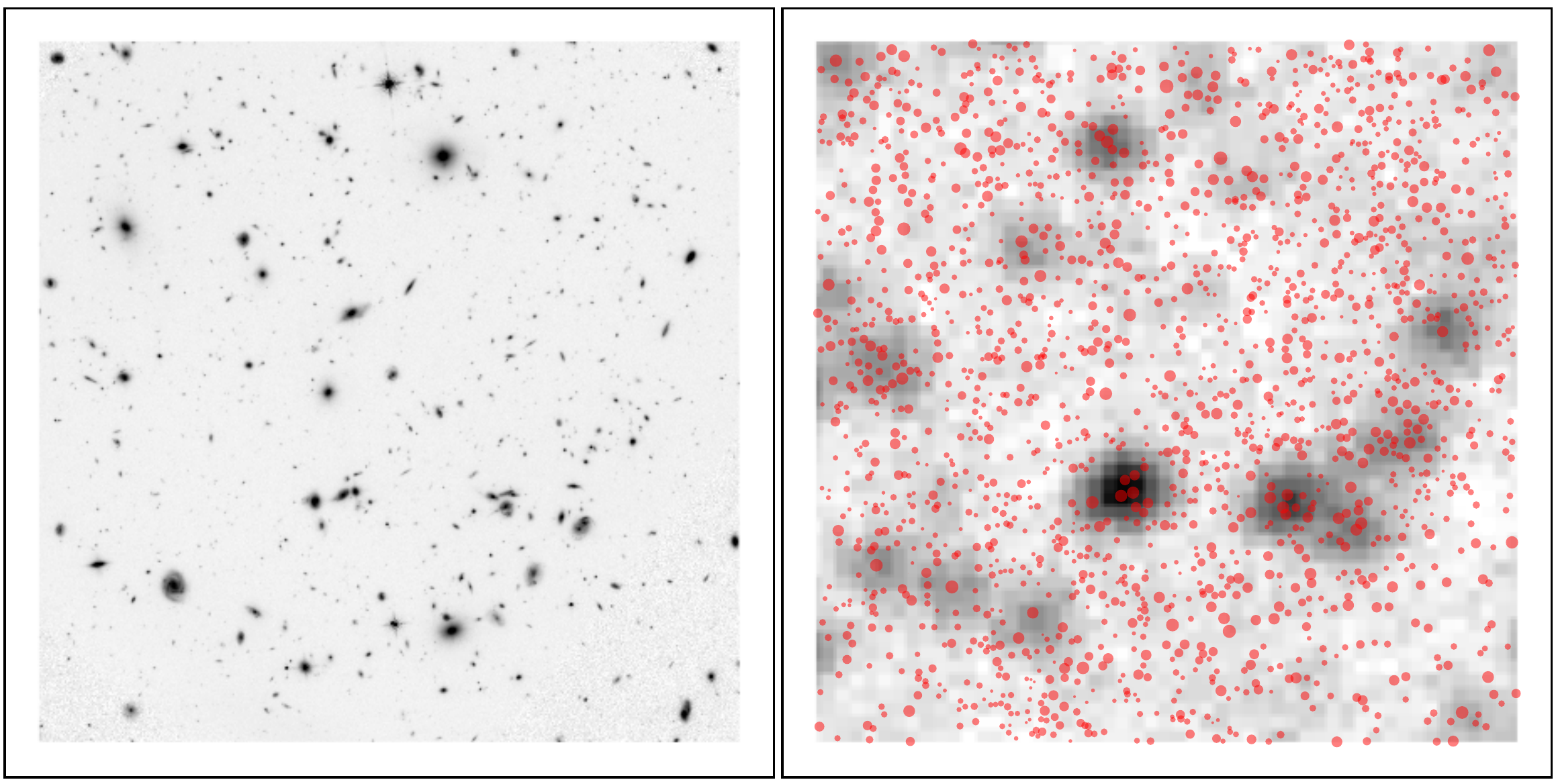}
\caption{ Illustration of the confusion problem. On the left is the \Hubble image of the Ultra-Deep Field at 1.6\micron. On the right is the deepest \Herschel PACS 160\micron image of the same field. The positions of the sources detected in the \Hubble image (red dots) are overlaid on the \Herschel image with their size proportional to their H band magnitude}.
 \label{fig:intro_fig}
\end{figure}

For deep extragalactic surveys, reality is somewhere in between these two idealized cases. The detected source density in the deepest PACS \citep{pacs} 160\micron images is $\sim3.1$ sources arcmin$^{-2}$ \citep{magnelli_13}. In the same region of the sky, the average detected source density in the shallowest tier of the \Hubble CANDELS \citep{Koekemoer:2011ApJS.197.36,Grogin:2011ApJS.197.35} 1.6$\mu m$ image is $\sim$200 arcmin$^{-2}$ and for the Hubble Ultra-Deep Field \citep[HUDF,][]{beckwith_06} is $\sim$1200 arcmin$^{-2}$. This is illustrated in Figure \ref{fig:intro_fig}.

From existing archival data, we have excellent estimates on the position, redshift (spectroscopic and photometric), UV--NIR SED shape, morphology, size and axis ratio of  sources that are {\it not} individually detected by {\it Herschel}, with which we can constrain their $L_{IR}$ \citep{wuyts_11,dale_07}. To date, none of this information has been used to help reduce the confusion in the \Herschel images. In this paper we show that provided we can constrain the $L_{IR}$ of the galaxies to within an order of magnitude using all the above properties, we can reduce the confusion noise significantly and obtain reliable photometry for much fainter objects.
\section{Method}\label{sec:method}

\begin{figure*}
\includegraphics[scale=1]{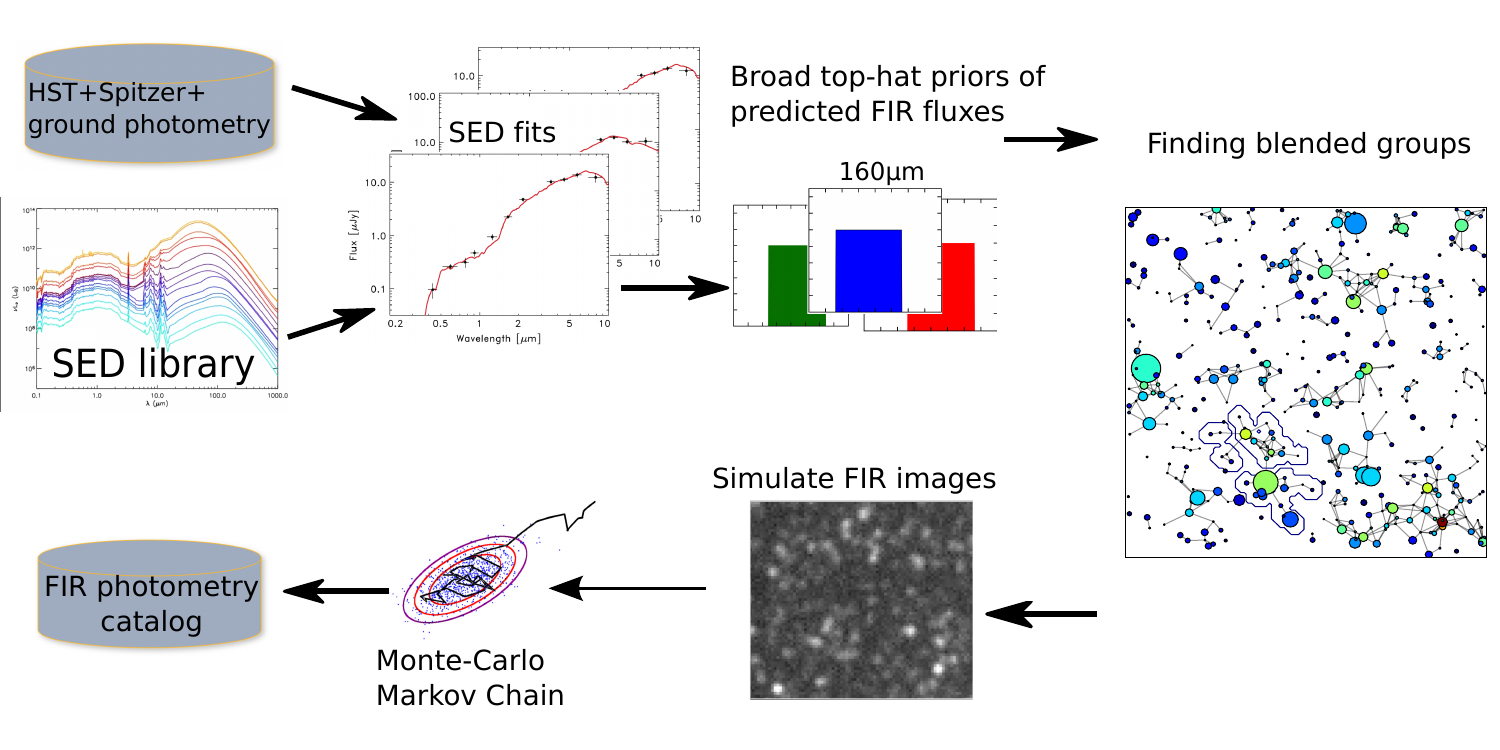}
\caption{Illustration of the fitting procedure. Semi-analytical models are used to generate an SED library based on realistic star-forming histories and chemical evolution and dust treatment. The SEDs are compared to the HST+\Spitzer+ground-based photometry. This comparison yields a probability distribution function (PDF) for predicted fluxes in each FIR band, which we broaden into top-hat PDFs centered on the peak of the PDF from the SED fit, illustrated in the top center of this diagram. The sources predicted to be brighter than some minimum flux are grouped together based on criteria that combine proximity and predicted brightness of the neighbors. Each circle shows one source with its size being proportional to the source's predicted flux. Each circle is color coded based on the number of neighbors it has. The contour plot around one of the disjoint graphs shows the pixels that are fit simultaneously for that group of sources. The sources in the disjoint graphs with an additional parameter accounting for the background are simultaneously fit and the posterior distributions are determined via a MCMC simulation.}\label{fig:fig1}
\end{figure*}

The procedure for improving the \Herschel photometry using Bayesian priors is outlined in Figure \ref{fig:fig1} using a PACS-160\micron image as an example. 
We need both positional and flux priors for every single source that is detected in the \Hubble image. The positions come from \Hubble H band imaging and the the flux priors from both analyzing the PACS-160\micron image and also the short wavelength ($\lambda<2$\micron) SED fitting of the sources that are detected in the \Hubble images using a mock library of SEDs(sections \ref{sec:library},\ref{sec:priors}). 
For each source, we only need a rough estimate of the PACS-160\micron flux to be within $\pm$1 dex of the corresponding true value.
We use our best-guess prediction for the PACS-160\micron fluxes (section \ref{sec:initial_guess}) of the sources along with
their positions to feed a graphical algorithm (section \ref{sec:find_blended_groups}) which
breaks the image into smaller regions each identifying the blended groups.
Subject to the positional and flux priors, a Markov Chain Monte Carlo (MCMC) simulation is run for each blended group with the number of dimensions being the number of sources with an additional background level to produce
an estimate for the full posterior distribution of the flux of each source (section \ref{sec:mcmc}).

\subsection{Library}\label{sec:library}
The library we propose to use for SED fitting and estimating the FIR flux of a given source is a set of simulated CANDELS light cones based on semi-analytic models \citep[SAMs;][Somerville et al 2014 in preparation]{somerville_08,somerville_12,fontano_09,lu_14}. The library should be large enough so that cosmic variance would not play a major role and using all of the CANDELS fields lightcones reduces the cosmic variance to the desired level. 
The SAM is applied on a set of halo merger trees extracted from a large cosmological $N$-body simulation which has a box size of 250$h^{-1}\,Mpc$ on a side with a mass resolution to follow galaxies with a stellar mass $\sim 10^8 M_{\odot}$. Following the merger trees, the model calculates the rates of gas cooling, star formation, outflow induced by star formation feedback, and galaxy-galaxy mergers. The model predicts realistic star formation histories and metallicity histories for (a number) galaxies in the entire volume in a cosmological context. These are used to produce SEDs based on the BC03 model for the library. The SED library includes normal star forming, quiescent and starburst galaxies.  In this set of SAM SEDs, FIR flux of galaxies are estimated via a slab model for dust attenuation and estimating the amount of absorbed starlight ($L_{IR}$) based on inclination of the model galaxy and its face on optical depth value. The FIR flux in each band is based on the shape of the FIR SED via templates of \citet{chary_elbaz_01} while other templates could be used as well.

\subsection{Priors}\label{sec:priors}
Our priors consist of strong positional priors and weak flux priors for the sources. Positions of the sources are taken from a deep, high-resolution image from the \Hubble Space Telescope. For the CANDELS fields, the detection
band will generally be at 1.6\micron, with a spatial resolution of
about 0.1 arc seconds. SExtractor \citep{sextractor} is used
for the source detection, and used for PSF-matched photometry of 
the sources in the \Hubble images. 
Our fitting technique requires priors for the FIR fluxes for every source detected in the \Hubble images. However, we find that these can be reasonably weak ($\pm$1 dex in flux) and still yield good photometry for most sources. We have verified (Safarzadeh et al 2014 in preparation) that we generally are able to predict the FIR fluxes of galaxies at low redshift to within $\pm$1 dex using only their SEDs at $\lambda<2$\micron for a sample of local normal star forming galaxies in SINGS sample \citep{kennicutt_03} and LIRGs in GOALS sample \citep{armus_09}. Apart from being able to estimate the \lirs for a given galaxy, a portion of the uncertainty for high redshift galaxies comes from the error in the photometric redshift estimates. Currently the uncertainty $\delta z /(1+z)$ is about 0.06 \citep{dahlen_13} in the CANDELS data which suggests that its effect on the SED fitting derived quantities will be negligible. However 5\% of the sources are outliers with the true redshifts significantly different from photo-z estimates. While the SED fits produce full probability distribution functions (PDFs) for the mid- and far-IR fluxes in each band, to avoid too much reliance on the SED models, we turn these PDFs into broad (2.2 dex wide) top-hat priors which are centered on the peak of each source's flux PDF.

\subsection{Initial source's flux estimate}\label{sec:initial_guess}
Our fitting procedure begins with an initial guess for the 160\micron flux for each source. These initial guesses are used to find the blended group of the sources which is the next step in our de-confusion method. For sources brighter than $3\sigma_{conf}$, the initial guess is based on the measured flux with the standard PSF matching photometry technique \citep{magnelli_13}. 

For sources fainter than $3\sigma_{conf}$, our initial guess will based on SED fitting. The SED for each source is fit using a library of SEDs we described in section \ref{sec:library}. The fits use photometry short-wards of rest-frame 2\micron-- i.e. light dominated by dust attenuated stellar emission.  Ideally, we would like to use as much
information as possible to make this prediction, including the spectroscopic
redshift, the full SED from UV to mid-IR, and possibly the galaxy 
axial ratio, size and morphology. For this proof-of-concept, we have
simply assumed that for every galaxy that is detected in \Hubble images, we can predict the \Herschel PACS-160\micron flux to within an order 
of magnitude. As described later, we expect that we can generally identify
post-facto the cases where this prediction has failed, and iterate the 
procedure to address this.

\subsection{Decomposing domains for image fitting using graphs}\label{sec:find_blended_groups}
In this method we use graphs to identify the most blended groups of sources in the PACS image. The graph is constructed on the PACS image but the sources (nodes) that makes up the graph are those sources that are detected in the \Hubble image. Each node is connected to other nodes in the PACS image if there is a strong interaction between them according to their predicted PACS-160\micron flux and distance from each other. Flux in a given PACS pixel at $\bf r$ from a source $x$ at position $r_{0}$ is:
\begin{equation}
\label{eq:graph_eq}
F^{\prime}_{\bf r}=F_{x,\bf r_0} * PSF(\bf r,\bf r_0) 
\end{equation}
if $F^{\prime}_{\bf r} > \alpha \times F_{y,\bf r}$,then sources $x$ and $y$ are connected with each other. For a PSF centered at $\bf r_{0}$, $PSF(\bf r,\bf r_{0})$ gives its value at position $\bf r$ (the PSF is normalized such as its central pixel value is 1) and * denotes the convolution operation. $\alpha$ is the sensitivity parameter. $\alpha = 1$ implies we only connect source A and source B if flux of source A at the position of source B is more than flux of source B at its central pixel position. $\alpha =0.1$ would imply connecting sources A and source B if flux of source A at the position of source B is more than 10\% of flux of source B at its own central pixel position (so more sensitivity). In this paper, we have used $\alpha=0.5$ (lower values of $\alpha$ will result into large graphs that expand the whole image whose analysis is computationally expensive). The result is a set of disjoint graphs that can be independently analyzed. We separately focus on each disjoint graph, select pixels that encompass that graph and constrain ourselves to those pixels when fitting the image.

\subsection{Fitting the image}\label{sec:mcmc}
\begin{figure*}
                \includegraphics[width=0.5\textwidth]{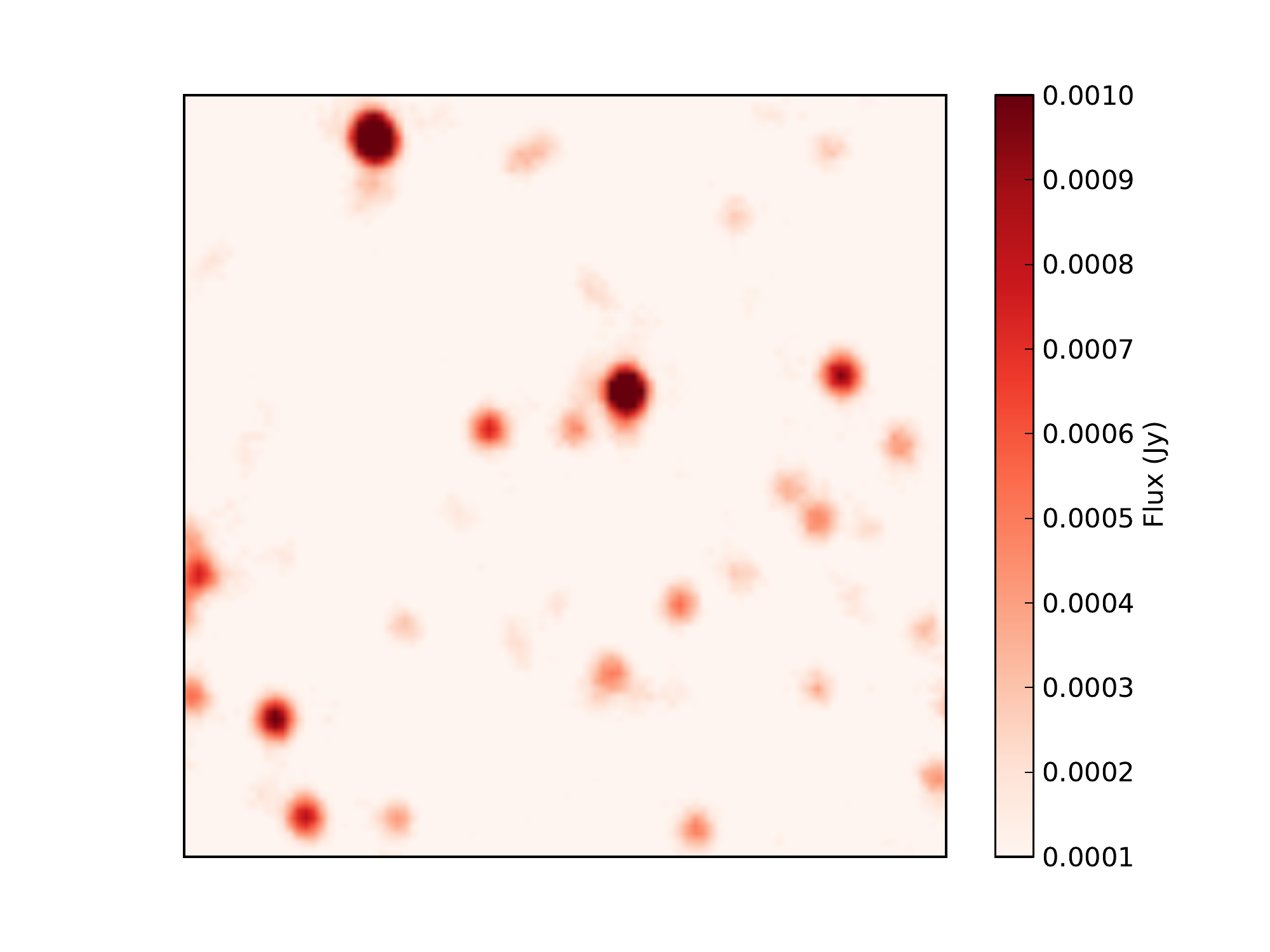}
                \label{fig:simulated_image}
                \includegraphics[width=0.5\textwidth]{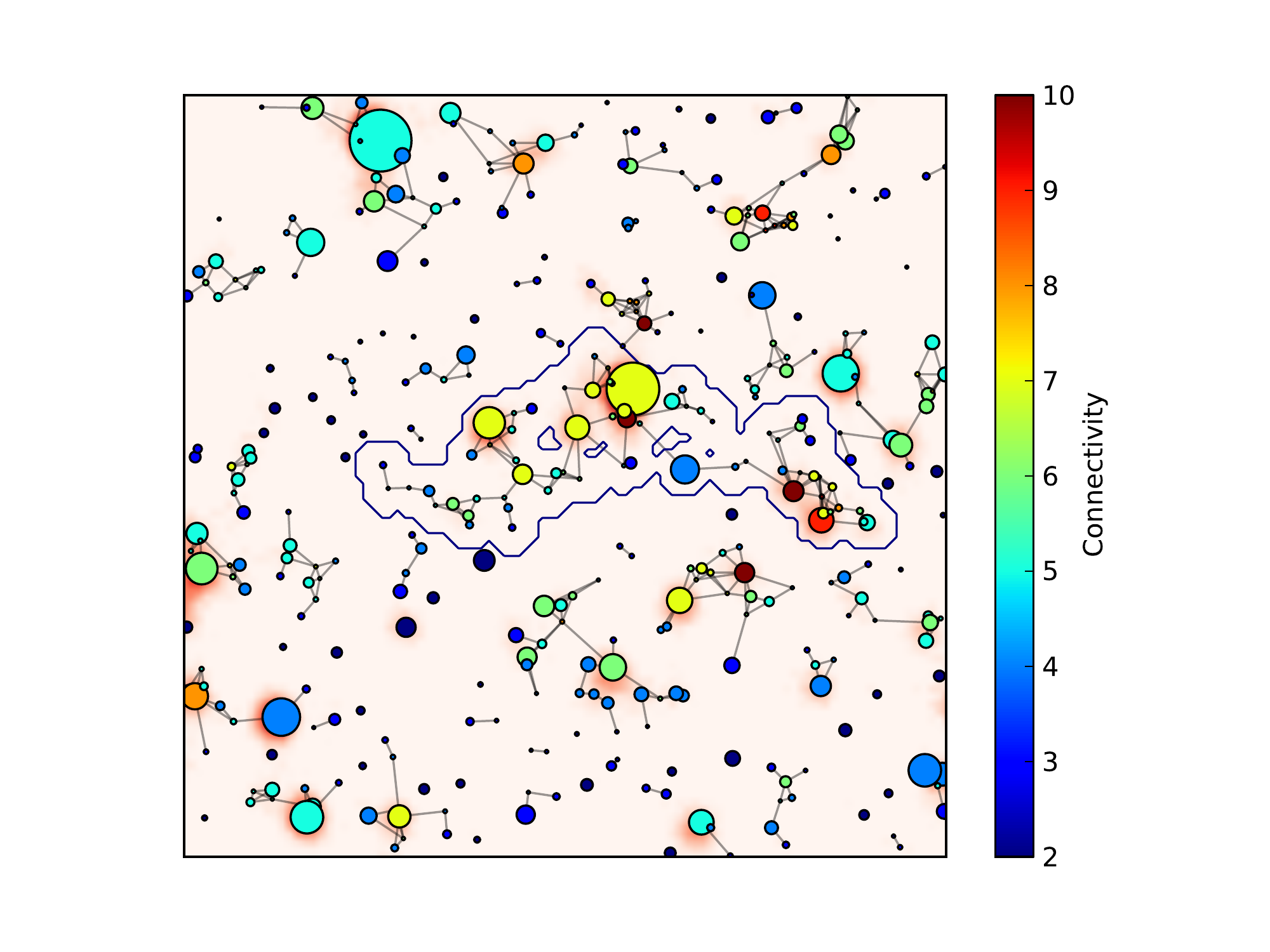}
                \label{fig:simulated_with_nodes}
        \caption{The process of fitting the simulated images. Panel (a) shows a section of the simulated PACS-160\micron image of GOODS-S field. Panel (b) shows the same image overlaid with the graph whose nodes represents the galaxies and color coded by the number of neighbors each node (galaxy) has. The size of each node is proportional to the galaxy's flux and in this image we use true flux for each node (fluxes are not dispersed). The presence of an edge between two nodes in the graph implies strong influence of one's flux on the other due to proximity and brightness. We have dispersed the flux of the sources that are not detected in the PSF matching analysis by 1 dex and then constructed the graphs. The contour plot around one of the disjoint graphs shows the pixels that we fit for in order to infer the flux of the blended sources in that graph. In panel (b), the color of each point indicates the number of neighbors of each node, as indicated on the color bar.}\label{fig:the_process}
\end{figure*}

Sources in a blended group have to be fit simultaneously. Due to the dispersion of the fluxes of the sources that are not detected by PSF matching technique, the resulting graphs have a different connectivity of nodes than the graph based on the true input fluxes, but this is what will happen in reality. Figure \ref{fig:the_process} shows the result of grouping the sources after dispersing their true fluxes by 1 dex. We draw a contour around one of the isolated graphs to show the pixels that will be analyzed in studying the sources in that particular blended group of sources.

We select a disjoint group of blended sources in the simulated PACS-160\micron image and estimate their flux through MCMC sampling. We chose to implement EMCEE, a python based affine invariant sampler for our purpose \citep{emcee_13}. The convergence of the chains are based on the Gelman-Rubin test \citep{Gelman1992} and requiring the $\chi^{2}_{r}=\chi^{2}/(N_{p}-N_{s}-1)$ to be close to 1. $N_{p}$ is the number of pixels we use to fit when estimating the flux of $N_{s}$ blended sources together. $\chi^{2}$ is defined as:
\begin{equation}
\label{eq:chi2_eq}
\chi^{2}=\sum_{j=1}^{N_{p}} \frac{  (\sum_{i=1}^{N_{s}}F_{i}*PSF_{i,j} - I_{j} )^{2} }{\sigma_{RMS}^{2}}
\end{equation}
where $F_{i}$ is the flux of source $i$ and $PSF_{i,j}$ gives the value of the PSF (centered on the position of source $i$) at the pixel $j$ and * denotes the convolution operation. $I_{j}$ is the value of pixel $j$ flux in the image. $\sigma_{RMS}$ is the instrumental pixel noise of the image which is the same as the science image.
The variables we fit for are flux ($F_{i}$) of $N_{s}$ sources with an additional background level flux that is not zero due to the presence of faint sources.

\section{Test of the method}\label{sec:test}
In this section we present a demonstration of the technique we proposed on a simulated PACS-160\micron image.

\subsection{Constructing simulated PACS-160\micron image}\label{sec:simulated_images}
\begin{figure*}
                \includegraphics[width=0.5\textwidth]{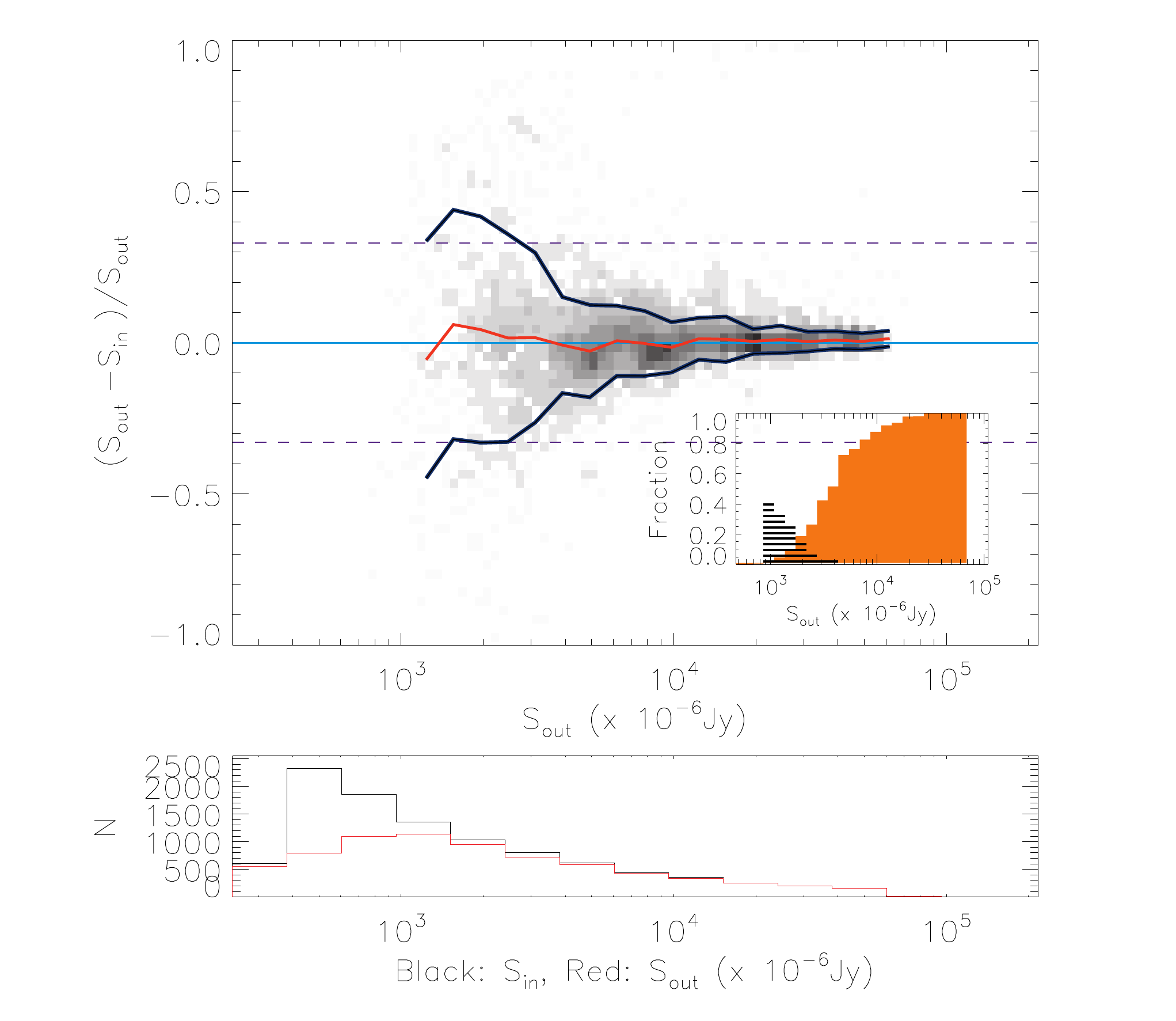}
                \label{fig:science_analysis}
                \includegraphics[width=0.5\textwidth]{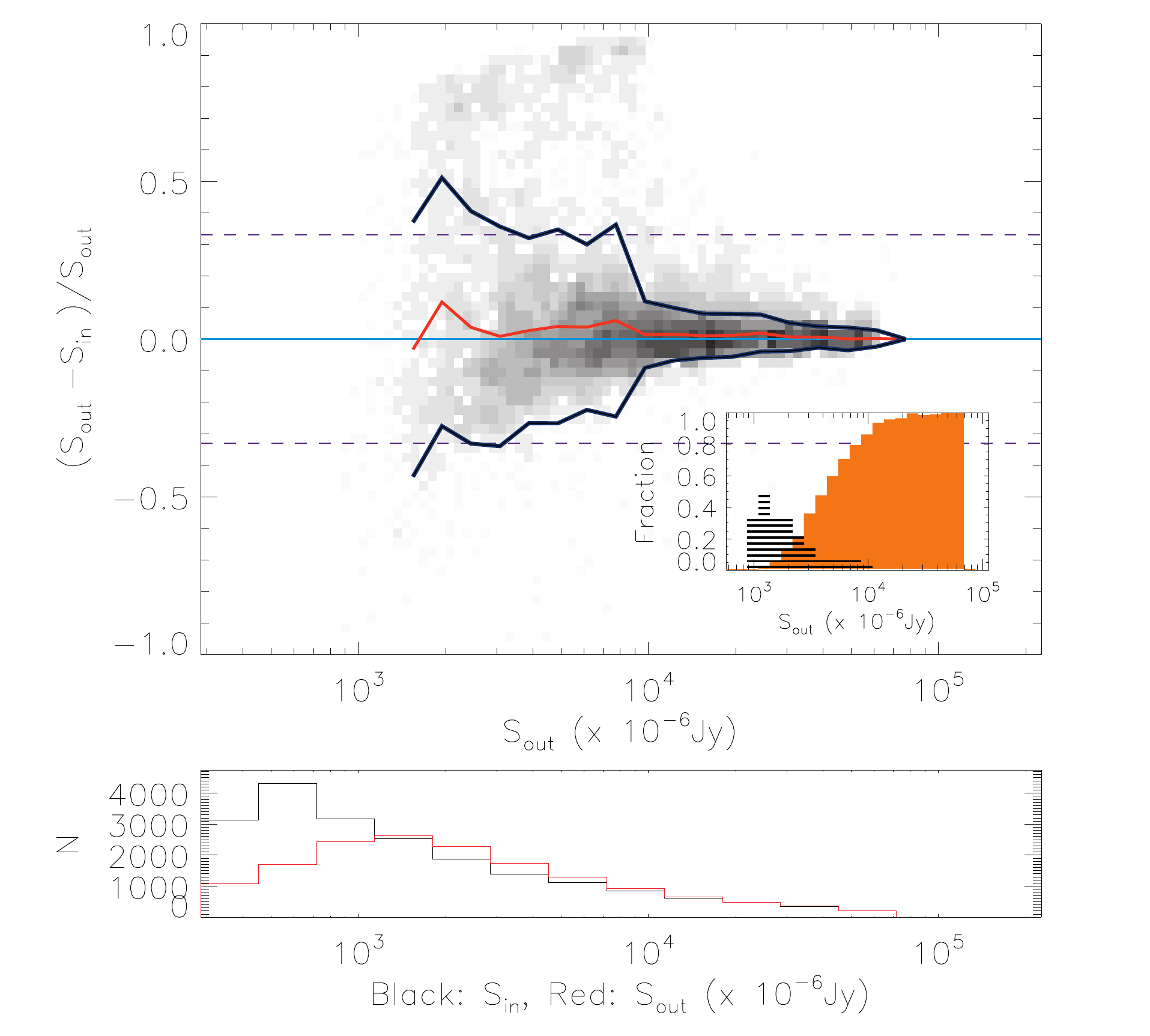}
                \label{fig:simulated_analysis}
        \caption{a)Results of MC simulations in the GOODS-South field using PSF matching method. Blue lines represent the average photometric accuracy defined as the standard deviation of the $(S_{out} - S_{in} )/S_{out} $ distribution in each flux bin (after 3$\sigma$ clipping). Red lines show the mean value of the $(S_{out} - S_{in} )/S_{out} $ distribution in each flux bin. Inset plots show the fraction of artificial sources detected in the image (i.e., completeness) as a function of input flux (orange plain histogram) and the fraction of spurious sources (i.e., contamination) as a function of flux density (striped black histogram). b)Results of MC simulations on our simulated PACS-160\micron image. The lines are the same as in (a). The science and simulated PACS-160\micron images are statistically similar.}\label{fig:comparing}
\end{figure*}

In order to account for the effect of clustering on the resulting
confusion noise, for our simulations we use the actual positions of
detected sources in the CANDELS GOODS-S catalog. 
For the purposes of our simulation, we need only a plausible SED 
from the optical through the FIR, not one that is necessarily close
to the truth for that galaxy. We use a custom-generated GOODS-South light cone mock catalog 
constructed from a SAM \citep{somerville_08,somerville_12,fontano_09}
to provide the library of physically plausible galaxy SEDs. For each
galaxy in the CANDELS image, we select galaxy at random from this
catalog within 0.5 mag in H-band apparent magnitude and 0.05 in
redshift. These are inserted as point sources--as the intrinsic size of the galaxies in GOODS-S PACS images is negligible--
and convolved with \Herschel PACS-160\micron PSF without adding a background level to the image.  
Noise is added to the image from the RMS map of PACS science image. Figure \ref{fig:comparing} illustrates the statistical similarity between our simulated image and the science image of PACS-160\micron of GOODS-South field. 

\subsection{Flux priors and initial guesses}
We analyze the simulated PACS-160\micron image with PSF matching photometric technique \citep{magnelli_13} and for sources which are detected to be brighter than $3\sigma_{conf}$, we use their measured flux as our initial guess. In order to do so, we simulated the corresponding \Spitzer MIPS 24\micron and \Spitzer IRAC Ch1 of our PACS-160\micron simulated image. In PSF matching technique, the MIPS and IRAC images are used as priors for analyzing the PACS image. The current procedure for source detection and construction of the current \Herschel catalogue is outlined in \citet{elbaz_11,magnelli_13}.

For sources fainter than $3\sigma_{conf}$ in the PACS image, because in our simulation we know the true flux of the sources, we disperse their flux by drawing from a uniform random deviate within a range $\pm$1 dex of the true value to mimic our ability in predicting the FIR fluxes to within an order of magnitude and use that as our initial guess. It should be noted that although we propose to use SAMs libraries for SED fitting and use the predicted FIR flux as the initial guess, for the demonstration of the method, we only disperse the fluxes to within $\pm$1 dex to mimic our inability to predict the precise FIR flux of the sources. If we were to use the SAMs library to predict the FIR flux via SED fitting, as our sources are drawn from the same library, the predictions would have been unrealistically close to the true values.

\subsection{Results}\label{sec:result}
\begin{figure}
 \includegraphics[scale=0.45]{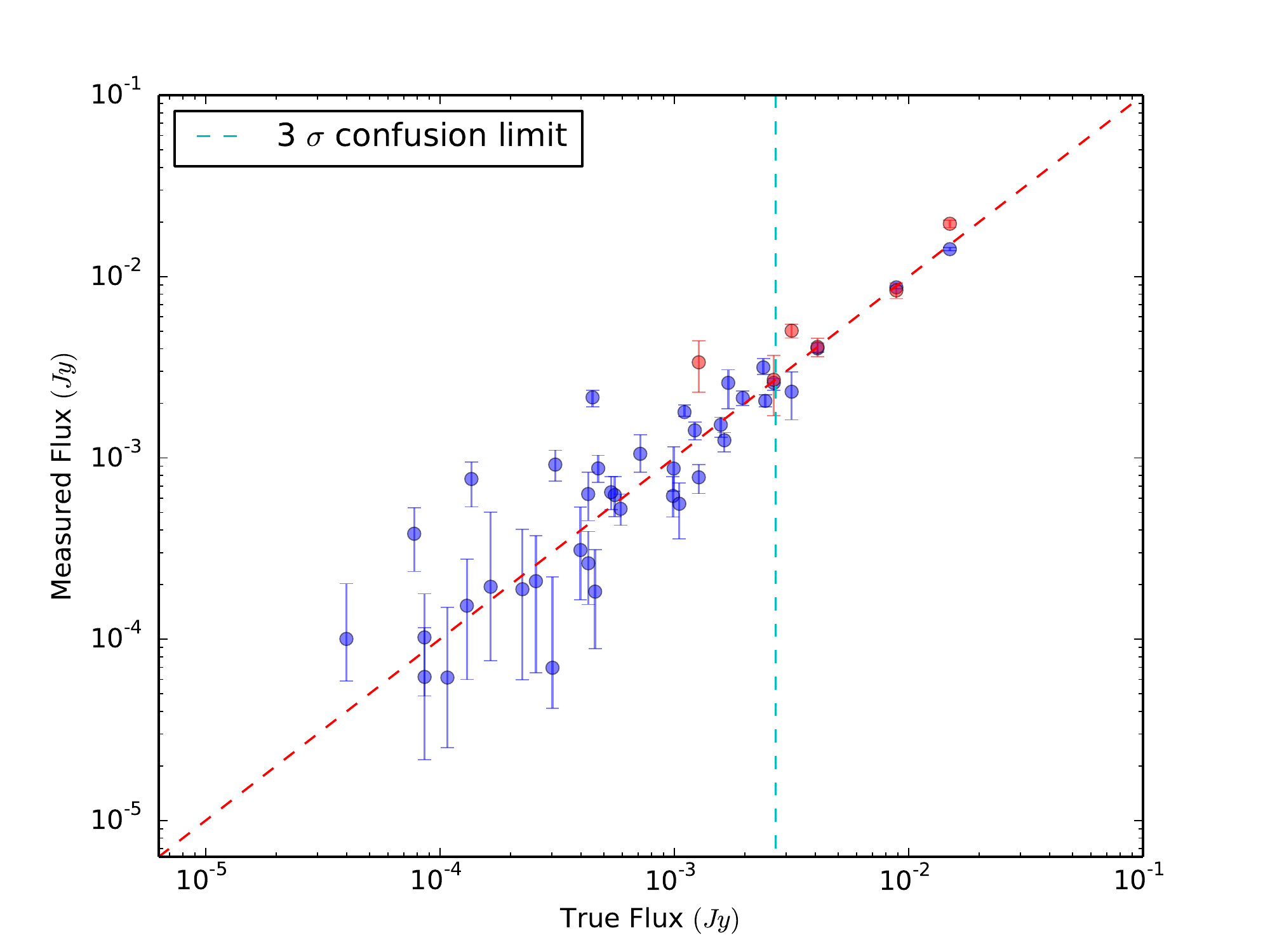}
\caption{ The measured flux from our method (Blue dots)  compared with the fluxes estimated from the more conventional technique of \citet{magnelli_13} (in red). Only about 10\% of the sources brighter than instrumental noise limit (0.2 mJy) are detected in \citep{magnelli_13}. The $3\sigma_{conf}$ is 2.7 mJy in our simulated image which is shown by the blue vertical line.\label{fig:1593result}}
\end{figure}

\begin{figure}
\includegraphics[scale=0.45]{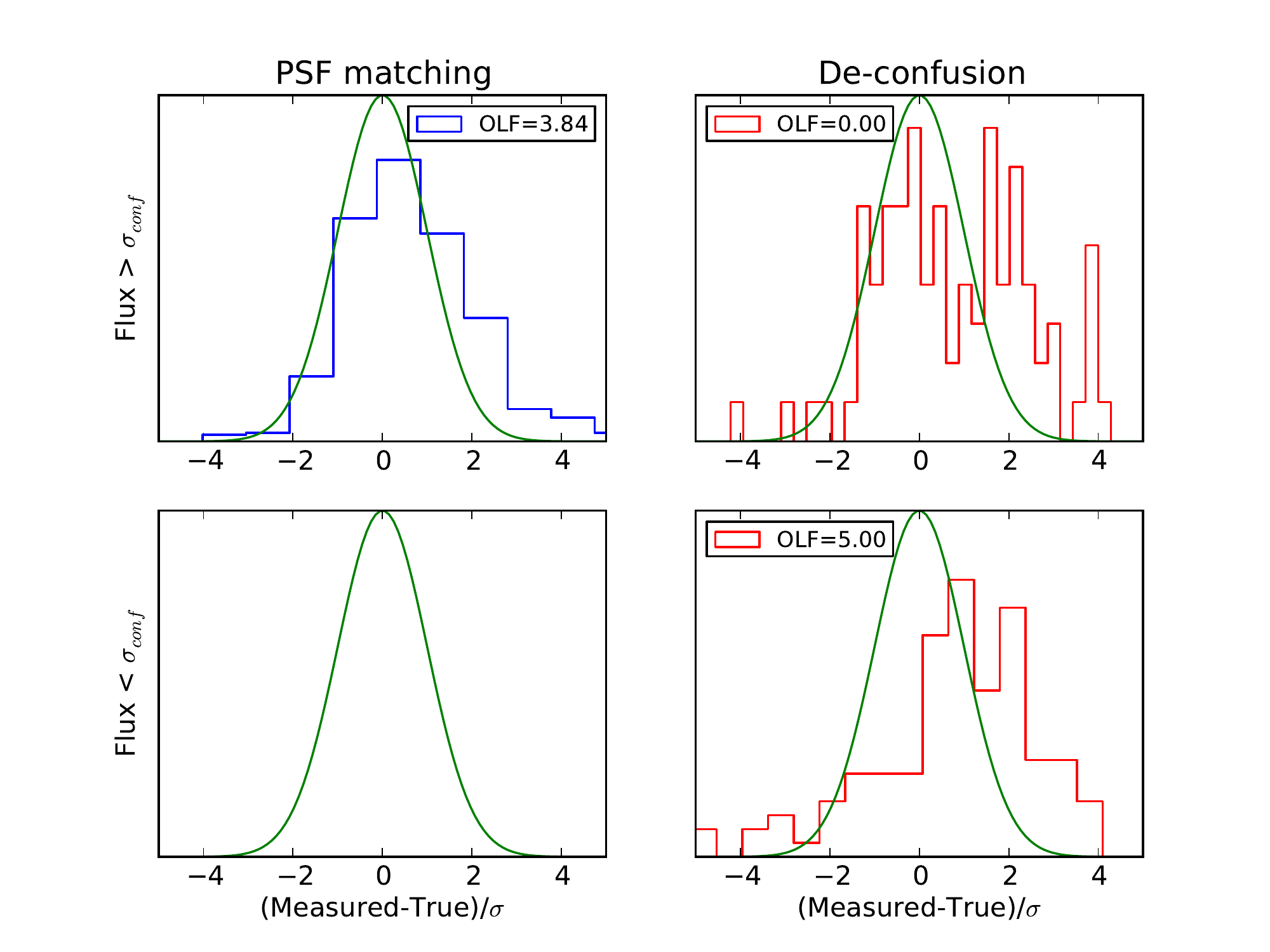}
\caption{Comparison of the percentage of the OLF (defined in the text) of de-confusion method and the standard photometry on a simulated PACS-160\micron image. The top row shows the result of flux measurement of sources that are detected to be brighter than the 1$\sigma_{conf}$ in the simulated image. The bottom row shows the result for the sources that are detected to be below the $1\sigma_{conf}$. There is no source detected below this limit in the standard photometry technique while our method only results in OLF of 5\%. Due to the computational costs of the MCMC, we have run our demonstration on only about a quarter of the image. Therefore the number of sources shown in the top panel is about four times larger for the PSF-matching technique (left) than for our de-confusion technique (right). In all the panels, the green is a Gaussian curve with mean zero and standard deviation of 1. The histograms would follow the green curve if sources were isolated (not blended) and the background noise is gaussian.}\label{fig:comparison}
\end{figure}

For a given blended group of sources that we analyze with our technique, we compare our result with the standard photometry technique \citep{magnelli_13,elbaz_11} on the same set of sources. The result is presented in Figure \ref{fig:1593result}. In the standard technique, sources with true flux fainter that $3\sigma_{conf}$ of 2.7 mJy are barely detected. For sources below this limit, only statistical upper limits are provided, based on the overall confusion noise of the image, not an upper limit based on the crowding of each individual source. Even for sources that are brighter than this limit, flux estimates from the standard technique to be biased high relative to the true fluxes, due to the contribution of nearby neighbors. The de-confusion method presented in this paper is able to probe sources as faint as the instrumental noise limit giving a posterior PDF for each individual source. The overall performance comparison of de-confusing technique and the standard technique is illustrated in Figure \ref{fig:comparison}. We define the outlier fraction (OLF) as the fraction of the sources whose measured flux is $5\sigma$ ($\sigma$ denoted the uncertainty on the measured flux) away from the true flux. For sources brighter than $3\sigma_{conf}$, the OLF is about 4\% in the standard method and 0\% in the de-confusion method. For fainter sources, while there is no detection in the standard technique, the OLF is about 5\% in de-confusion technique. 

 \section{Summary and Discussions}\label{summary}
 We have demonstrated via simulations that we are able to obtain
reliable fluxes for sources significantly below the nominal confusion
limit of the deepest 160\micron \Herschel images. To achieve 
this, we use strong priors on source positions and weak top-hat priors of 
$\pm 1$ dex wide on source fluxes. We also assume perfect knowledge
of the point-spread function. 

In a confused image, it is crucial to simultaneously estimate the flux of sources that are strongly affect by each other due to their proximity in position and relative brightness. In order to do so, we have developed a graphical method for identifying sources that need to be 
fit simultaneously. This also makes the problem computationally tractable as we do not have to fit for all the HST sources in the image at once. This segregation results in imperfect
estimates of photometry and uncertainties for some sources. Nevertheless,
our simulations indicate that both our fluxes and our uncertainties
appear to be generally reliable down to 1 mJy, which is a factor
of $\sim3$ below the nominal $3\sigma_{conf}$ confusion limit. 

In the case of real data where we do not know which sources have good photometry in our technique, there are various tests that can be carried out to identify possible problems. 

(1) Analyzing the residual image: Comparing the result of the image based on the predicted fluxes with the original image can reveal if there was a source not included in the set of priors or in the group that was simultaneously de-blended. This could also be revealed in the $\chi_r^2$ value of the fits as well.

(2) Repeat the flux estimation within a different graph. Using different initial guesses
for the fluxes of sources will result in different connectivity between sources, and hence
different blended groups of sources that will be simultaneously fit. Repeating the photometry with
a variety of initial guesses will then provide multiple flux estimates for each source, and
sources with a wide range of predicted fluxes can be more easily identified and flagged.

(3) Decreasing $\alpha$. The parameter $\alpha$ which is described in section \ref{sec:find_blended_groups}, determines what sources are considered to be in a given blended group. Decreasing $\alpha$ will result in having more sources joining a given blended group and increases the number of sources that have to be fit simultaneously. Although higher number of sources to fit simultaneously will be more computationally expensive, the resulting photometry is more accurate. If there is a faint source whose flux is easily affected by other far distant sources, decreasing $\alpha$ to values around 0.1 or even less could help improve the photometry. 

(4) We expect that there will be a few optically-undetected Herschel sources in each CANDELS field. This will lead to posterior PDFs for a few groups of sources that are catastrophically wrong. Cases like this can be revealed by comparing the initial guesses to the measured fluxes, and examining cases where the fluxes moved unusually far (and generally became brighter) than the initial guesses.

\section{Future improvements}\label{sec:future}

While we consider the method we have outlined so far to be robust enough
already to construct a good photometric catalog, there are various ways in which
it can be improved and extended. We briefly outline these here.

(1) Improving the priors. We show in a companion paper (Safarzadeh et al 2014 in preparation) that we can typically predict FIR fluxes of low-redshift galaxies to within $\pm$1 dex using existing SAM SED libraries. Compared to local star-forming galaxies, main-sequence galaxies \citep{noeske_07} at high redshift have higher SFR \citep{whitaker_12}, higher gas fractions (\citealt{magdis_12,daddi_10}), smaller sizes \citep{vanderwel_14}, different morphologies \citep{dekel_09} and chemistry \citep{magdis_10}. SAMs have been successful at generating the same trend for high redshift galaxies and we intend to use SAM libraries for predicting the FIR flux of high redshift galaxies. For the first-generation catalog, we envision using the results of SED fitting of the photometry short wards of rest-frame 2\micron to predict the FIR fluxes. However, we are well aware that this is not ideal and there is room for improvement by considering other information such as galaxies size, axial ratio and morphology. 

(2) Fit multiple bands. The demonstration here used only the \Herschel PACS-160\micron image (although we used the simulated IRAC 3.6\micron and MIPS 24\micron images to drive the standard PSF-fitting photometry for the simulation). Our goal is to apply the de-confusion technique
to all the FIR bands in order to weed out sources with discrepant photometry. For example, if the measured observed 160\micron flux is significantly different from the measured 100\micron flux, the measured photometry should be flagged. The most important bands to include are those in SPIRE, which are heavily confused for most of the sources of interest at $z>2$. 

(3)De-confusing the PACS-100\micron and 160\micron images will give us the rest frame 40\micron and $\sim$ 50\micron for a $z\sim3$ galaxy, which is not very close to the peak of the FIR SED. However, de-confusing the SPIRE \citep{spire} images -- which trace the peak of FIR emission for galaxies at $z\sim3$ --  can better constrain the FIR luminosity.  In order to fit the SPIRE images, we plan to use the result of de-confusing PACS-100\micron and 160\micron and together with shorter wavelength SED priors, de-confusing SPIRE images, starting with SPIRE 250\micron and progressively move up to de-confusing longer wavelength SPIRE images.

\section{Acknowledgements}
This research was partially supported by HST Program GO-12060 provided by NASA through grants from the Space Telescope Science Institute, which is operated by the Association of Universities for Research in Astronomy, Inc., under NASA contract NAS5-26555. This work was also partially supported by NASA through an award issued by JPL/Caltech. We are grateful to Mark Dickinson for help and advice on this project.

\bibliography{apjmnemonic,modified_version}

\end{document}